# Quantum-size effect induced Andreev bound states in ultrathin metallic islands proximitized by a superconductor

Guanyong Wang,[1,2,†] Li-Shuo Liu,[3,†] Zhen Zhu,[1] Yue Zheng,[3] Bo Yang,[1] Dandan Guan,[1] Shiyong Wang,[1] Yaoyi Li,[1] Canhua Liu,[1] Wei Chen,[3] Hao Zheng,[1,*] Jinfeng Jia[1,4,5,*]

1. Tsung-Dao Lee Institute, Key Laboratory of Artificial Structures and Quantum Control (Ministry of Education), School of Physics and Astronomy, Shanghai Jiao Tong University, Shanghai 200240, China

2. International Quantum Academy, Shenzhen 518048, China

3. National Laboratory of Solid State Microstructures, School of Physics, and Collaborative Innovation Center of Advanced Microstructures, Nanjing University, Nanjing 210093, China

4. State Key Laboratory of Quantum Functional Materials, Department of Physics, and Guangdong Basic Research Center of Excellence for Quantum Science, Southern University of Science and Technology, Shenzhen 518055, China

5. Quantum Science Center of Guangdong-Hong Kong-Macao Greater Bay Area (Guangdong), Shenzhen 518045, China

ABSTRACT

While Andreev bound states (ABSs) have been realized in engineered superconducting junctions, their direct observation in normal metal/superconductor heterostructures-enabled by quantum confinement-remains experimentally elusive. Here, we report the detection of ABSs in ultrathin metallic islands (Bi, Ag, and SnTe) grown on the $s$-wave superconductor NbN. Using high-resolution scanning tunneling microscopy and spectroscopy, we clearly reveal in-gap ABSs with energies symmetric about the Fermi level. While the energies of these states show no position dependence, their wave functions exhibit spatial oscillations, demonstrating a quantum size effect. Both the energy levels and spatial distribution of the ABSs can be reproduced by our effective model in which a metallic island is coupled to the superconducting substrate via the proximity effect. We demonstrate that the coupling strength plays a critical role in determining the ABS energies. Our work introduces a novel physical platform for implementing ABSs, which hold promise for significant device applications.

[†]These authors contributed equally to this work.
*haozheng1@sjtu.edu.cn; jfjia@sjtu.edu.cn

Andreev bound states (ABSs) emerge at the interface between normal metals (N) and superconductors (S) as discrete in-gap quasiparticles, arising from coherent electron-hole combinations via Andreev reflection [1-3]. These states are not only fundamental to understanding proximity-induced superconductivity but also hold transformative potential for quantum computing and low-dissipation electronics [4-8]. While ABSs have been extensively studied in engineered devices such as SNS nanowires or quantum dots (QDs) coupled to superconductors [9-17], their unambiguous observation in NS heterostructures has remained a longstanding challenge, despite tremendous efforts has been devoted [18-20]. Nanoscale metallic island/superconductor heterostructures are good candidates for the investigation of ABS. It is known that when the size of an island is reduced to the scale of its Fermi wavelength, pronounced quantum size effect will appear [21-26]. Quantized energy levels in 1D or 2D-confined systems have been widely reported, which exhibits hundred-meV-level spacing [27-31]. For 3D-confined electrons (*e.g.*, tens-nm-scale ultrathin islands), spacing collapses to a few meV – comparable to conventional superconducting gaps. In such circumstance, discrete ABSs may arise when the quantum levels are proximity coupled to a superconductor. However, such a quantum size effect induced ABS in NS heterostructure with confined-geometry has rarely been explored.

Here we present that quantum-size effect may provide alternative way to realize discrete in-gap ABSs in laterally confined ultrathin islands coupled to a superconductor, as well as induce a spatial modulation of the wavefunctions of ABSs. We prepared ultrathin metallic (Bi, Ag and SnTe) nanoislands on NbN substrate. Our observation of ABSs in simple and easy-fabricated structure may stimulate new device development based on NbN heterostructures, especially in Andreev qubit [4-7].

Experiments were conducted in a scanning tunneling microscopy (STM) system with a base pressure $<1 \times 10^{-10}$ Torr. Superconducting NbN films were grown on $SrTiO_3$(111) substrates using plasma-assisted molecular beam epitaxy [32]. Bi islands were thermally evaporated onto NbN and $NbSe_2$ substrates at room temperature, while Ag and SnTe were evaporated onto NbN at ~100°C. Samples were transferred *in situ* to the STM stage and cooled to 0.4 K for measurement using tungsten tips. Scanning tunneling spectroscopy (STS) data were acquired via lock-in technique (991 Hz, 0.05 mV modulation).

NbN is an isotropic s-wave superconductor with high critical temperature (17 K) and is a widely used substrate [33-35], but STM studies of NbN-based NS heterostructures are uncommon [36]. One of the major challenges is obtaining sharp interfaces and atomically flat surfaces in these samples. Here, we successfully grow single crystalline NbN films [32]. As displayed in Fig. 1(a), our NbN thin film with a thickness of approximately 20 nm shows atomically flat terraces. STSs (Figs. 1(f) and S1 in Supplemental Material [37]) exhibit fully-gapped "U" shapes with coherence peaks at $\pm2.30$ meV and without detectable spatial variations. It thus confirms the uniform s-wave superconductivity in our NbN thin films. Unlike the droplet shape, which is common when metals are deposited on a compound surface, our Bi islands have flat top surfaces parallel to the NbN substrate (Fig. 1(b)). The atomically-resolved STM image in Fig. 1(b) inset shows a distorted orthorhombic atomic lattice. When comparing shapes, height, and atomic structure of our islands to the previous results [43-45], one can know that the top surface is (110). In the following, we focus on the islands with height of ~0.9 nm, namely 1 bilayer (BL) Bi(110).

A STM image of 1-BL Bi(110) island is shown in Fig. 1(c). STS measured on NbN substrate (near the island) yields a fully-gapped spectral shape with no in-gap state (IGS) [black curve in Fig. 1(d)]. In contrast, STSs measured on the island reveal prominent discrete peaks within the superconducting gap. The peak energies are symmetric with respect to the Fermi level, but their intensities are asymmetric. We then employed spatially resolved STSs to investigate the flat top Bi islands. The results are shown in Figs. 2 and S2-S4 [37]. On 1BL-Bi(110) islands with different sizes, one, two, and multiple pairs of peaks may appear within the superconducting gap. As one can see, IGSs are present across the entire Bi islands (red curves in Fig. 2(b, e, f)), but absent on NbN substrate even very close to the Bi islands (black curves in Fig. 2(b, e, f)). Notably, the energy of IGS exhibits spatial homogeneity across one entire island.

In order to elucidate the origin of observed IGS, we carried out an effective model calculation. The model considerers a QD coupled to a superconductor. The Hamiltonian is given by [46],

$$H = \sum_\sigma \epsilon_{QD} d_\sigma^\dagger d_\sigma + \sum_{\vec{k},\sigma} \varepsilon_{\vec{k},\sigma} c_{\vec{k},\sigma}^\dagger c_{\vec{k},\sigma} + \sum_{\vec{k}} \Delta(c_{\vec{k},\uparrow}^\dagger c_{-\vec{k},\downarrow}^\dagger + c_{-\vec{k},\downarrow} c_{\vec{k},\uparrow}) + \sum_{\vec{k},\sigma}(t\, d_\sigma^\dagger c_{\vec{k},\sigma} + t^* c_{\vec{k},\sigma}^\dagger d_\sigma). \quad (1)$$

Here, $\epsilon_{QD}$ is the energy levels of the QD; $\varepsilon_{\vec{k},\sigma}$ is the normal electron dispersion of the superconductor. In the subsequent calculations, we assume a spin degeneracy and apply the free electron approximation. $\Delta$ is the pairing potential of the s-wave superconducting substrate; $t$ is the tunneling energy between the electronic states in the superconductor and the QD, which is assumed to be a real constant independent of the wave vector and spin. Within the Dyson equation, we obtain the Green's function $G_{QD}(\omega)$ of the QD [47]:

$$G_{QD}(\omega) = \left(g_{0,QD}^{-1} - \Sigma_{SC}(\omega)\right)^{-1}$$

$$g_{0,QD}^{-1} = \omega - \left(\frac{(\hbar\pi)^2}{2m^*}\left(\left(\frac{n_x}{L_x}\right)^2 + \left(\frac{n_y}{L_y}\right)^2\right) - \mu\right)\sigma_z$$

$$\Sigma_{SC}(\omega) = \frac{-\Gamma}{\sqrt{\Delta^2 - \omega^2}}\begin{pmatrix}\omega & \Delta \\ \Delta & \omega\end{pmatrix}, \quad (2)$$

$g_{0,QD}$ is the Green's function of the QD without coupling to a superconductor. We model the QD as a free electron (with effective mass m* and chemical potential μ) confined in a two-dimensional rectangular box with length of $L_x$ and $L_y$. $\Sigma_{SC}(\omega)$ is the self energy correction of the QD induced by Andreev reflection when it is coupled to a superconductor. The linewidth function of the superconducting substrate is given by $\Gamma = \pi\rho(k_F)t^2$, where ρ is the density of states (DOS) on Fermi energy. Note that Γ also quantifies the coupling strength between the QD states and the superconductor. From Eq. 2, we know that incorporating a superconducting substrate to the QD states will not only mix electron and hole states but also lead to an energy correction to the original QD states. Particularly, when $|\omega| > \Delta$ the diagonal terms of $\Sigma_{SC}(\omega)$ are purely imaginary, which induces a broadening of the energy levels in the QD. When $|\omega| < \Delta$, the diagonal terms of $\Sigma_{SC}(\omega)$ becomes real, which only shift the energy level. The off-diagonal terms of $\Sigma_{SC}(\omega)$ mixes the electron and hole part, which will duplicate the energy levels, *e.g.* establish a pair of ABSs. Moreover, if the coupling strength $\Gamma$ is appropriate, discrete quantum levels above the superconducting gap will be smeared out due to broadening, while the levels within the gap will still give a sufficiently strong spectral weight.

Figure 3(a) displays the calculated the DOS of island #1 shown in Fig. 2(a) (The parameters used in the simulation are listed in Table SI [37]). We find the coupling strength $\Gamma$ plays a critical role in determining the IGS energy. In Fig. 3(a), the black curve marked as $\Gamma = 0$ depicts the simulated DOS of the island #1 without coupling to superconductor. Three discrete peaks near Fermi energy can be found. The level with quantum numbers $[n_x, n_y] = [2, 4]$ is located close to the zero energy and of particular interest. Upon turning on the coupling, the level evolves into one pair of in-gap ABS states, whose energies are symmetric with respect to the zero energy. It implies the formation of Bogoliubov quasiparticles through electron-hole mixing. With an increasing of $\Gamma$, the ABSs are pushed to higher energy. In the strong coupling regime ($\Gamma/\Delta \gg 1$), all ABSs merge into the coherence peaks, as shown in Fig. 3(a) (the dark green curve). This case becomes the conventional proximity effect induced "U-shaped" superconducting gap. We note that the simulated DOS at $\Gamma = 0.13\Delta$, in corresponding to the weakly coupling regime, fits well to our experimental data. (The states with quantum numbers other than $[n_x, n_y] = [2,4]$, *i.e.* [3,1] and [1,5] states, do not appear as discrete ABSs, but manifest themselves as broadened spectral signatures at energies outside the superconducting gap (see Supplemental Material Fig. S7(a) [37]). It is because their higher initial energies *i. e.* lager than the superconducting gap, require larger $\Gamma$ to generate observable in-gap spectral weight (see Supplemental Material Fig. S7(b) [37]).)

When the island size is only a few Fermi wavelengths, the electron forms a standing wave in real space due to quantum interference between incident and reflected waves at the boundaries. Because of the particle-hole symmetry in a superconducting system, the hole component shares the same periodicity as the electron component. Bogoliubov quasiparticle standing waves thus can be established. We now pay attention to the spatial distribution of the ABSs. Figure 3(b) presents a position-dependent STSs measured on island #1. One pair of ABSs with energies of $\pm$ 0.25 mV are resolved on the entire island. While the energies of the ABSs do not show position dependence, the intensities of the electron-like and hole-like states oscillate periodically in real space. Furthermore, figure 3(c) plots the experimental intensity profiles of the ABSs (black curve), revealing four anti-nodes consistent with the quantum number $n_y$=4. The simulated spatial intensity map of the ABS (Fig. 3(d)), based on a 2D rectangular quantum well model (see Supplemental Material for simulation details [37]) with parameters $[n_x, n_y] = [2,4]$ and $\Gamma=0.13\Delta$, reproduces the key features of the standing wave pattern, *e. g.* the energy of ABSs, the node/anti-node numbers, and the oscillation periodicity. Nevertheless, we also note some imperfects in the simulations. For instance, the experimental maxima exhibit slight variations in amplitude as comparing to the simulation (Fig. 3(c)). The possible reason is that we model the islands as ideal rectangular boxes, whereas real islands exhibit geometric distortions, *e. g.* non-rectangular shapes and rounded edges, which slightly alter the interference condition and the standing wave pattern. Despite this minor inconsistency, the agreement on essential characteristics supports the quantum confinement origin for the ABSs. Future work incorporating realistic island geometries and defect distributions may refine the spatial modulation details.

As a control study, we grew Bi nanoislands on $NbSe_2$. In Figs. 4(a)-4(c), STSs measured on $NbSe_2$ substrate and Bi islands both exhibit fully gapped spectra without IGS, which consisted with previous experiments of Bi and Ag deposited on $NbSe_2$ or Nb [48-51], where no IGSs are revealed. We conducted a theoretical fitting to the Bi/$NbSe_2$ data and deduced a coupling $\Gamma/\Delta$ of ~14 (see Supplemental Material Fig. S10 [37]), which is about two orders of magnitude larger than Bi/NbN. Additionally, we grew Ag and SnTe islands on NbN and clearly discern IGSs (ABSs) (Figs. 4(d)-4(f), S8 and S9 [37]). For example, two pairs of ABSs, whose intensities spatially oscillate, can be found in one Ag island on NbN (Fig. 4 (f)). This indicates that the weakly coupling regime consistently appears in NbN-based NS heterostructure. While the microscopic origin determining the coupling $\Gamma/\Delta$ is waiting for theorical investigation, we find it reconciles with interface quality. Epitaxy of Bi on Van der Waals material $NbSe_2$ facilities a low-defects interface (Fig. 4(a) inset), which may enhance the tunneling rate (t) and thus coupling strength $\Gamma$ ($\propto t^2$). In contrast, the significant lattice mismatch between NbN(111) (a=0.31 nm) and Bi(110) (a=0.45 nm) may induce interfacial disorder (Fig. 1(d) inset), which suppresses t and $\Gamma$. On the other hand, we notice a recent elegant experiment observing ABS in artificial QDs on Ag proximitized to Nb [52]. The weak coupling is achieved by separating the Ag surface states from the Nb with a thick Ag bulk state (~12 nm). However, when the thickness of Ag island is reduced to double layers, the ABSs disappear, because $\Gamma$ between Nb and Ag island is naturally large. It is also known that NbN ($NbSe_2$ and Nb) has a weak (strong) superconducting proximity effect [36, 51, 53]. We thus expect that a conventional superconductor with weak proximity effect may facilitate the observation of discrete ABSs in its NS heterostructure.

In summary, we directly visualized quantum-size-effect induced ABSs in ultrathin metallic islands proximitized by an s-wave NbN superconductor. We find the coupling strength plays a critical role in the appearance of ABS, and the intrinsic weak coupling strength of NbN enables direct observing of discrete ABSs in simple SN heterostructure. Key features of these ABSs, which are unambiguously linked to quantum size effects, are corroborated by our effective QD-superconductor model. This work not only resolves the long-standing challenge of realizing discrete ABSs states in simple NS systems but also deep our understanding of superconducting proximity effects.

*Acknowledgments*
We thank NSFC (Grants No.12488101, No. 92365302, No. 12104292, No. 12474156, No. 22325203, No. 92265105, No. 12074247, No. 12174252, No. 52102336), the Science and Technology Commission of Shanghai Municipality (Grants No. 2019SHZDZX01, No. 20QA1405100, No.24LZ1401000), the Ministry of Science and Technology of China (Grants No. 2020YFA0309000), Cultivation Project of Shanghai Research Center for Quantum Sciences (Grant No. LZPY2024-04) and innovation program for Quantum Science and Technology (Grant No. 2021ZD0302500) for financial support.

*Data availability* – The data that support the findings of this article are not publicly availability. The data are available upon reasonable request from the authors.

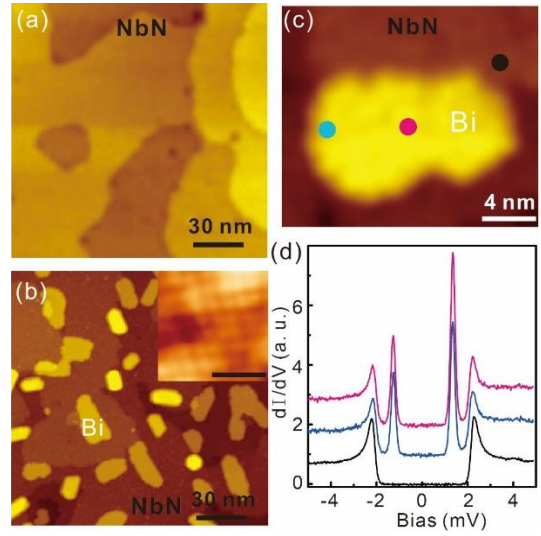

FIG. 1. (a) STM image of the NbN thin film. (b) Topography of Bi islands on NbN. The inset: an atom-resolved image of one Bi island, which is consistent with Bi(110). The scale bar is 1.5 nm. (c) A zoom-in image of a 1BL Bi(110) island on NbN. (d) Tunneling spectra measured on NbN (black curve) and on the Bi island. Blue, and red curves are obtained on the points with the same colors in (c) at temperature of 0.4 K.

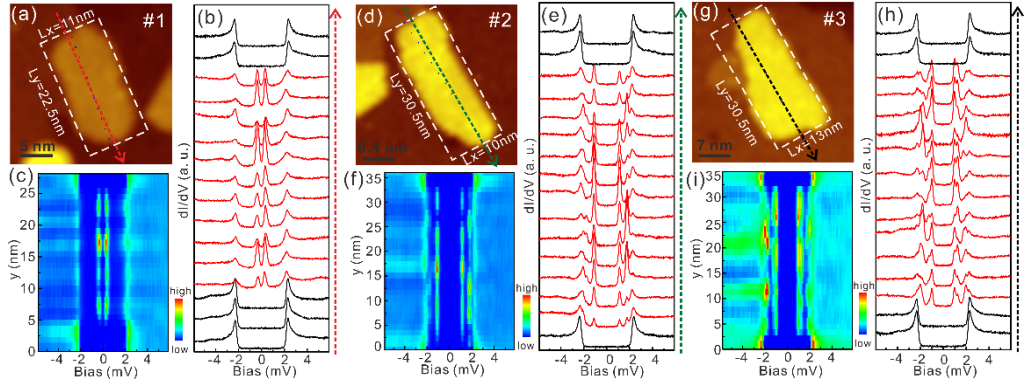

FIG. 2. (a, d, g) Topographies of 1BL Bi islands labeled as #1, #2 and #3, which lengths ($L_x$ and $L_y$) are indicated, on NbN. (b, e, h) Spatial profiles of the tunneling spectra (with offsets) measured along the arrows marked in (a, d, g). The black and red curves in (b, e, h) represent the spectra measured on NbN substrate and Bi islands respectively. Note one, two and multiple pairs of in-gaps states are resolved in (b, e, h). (c, f, i) Color plots of the position-dependent tunneling spectra shown in panel (b, e, h).

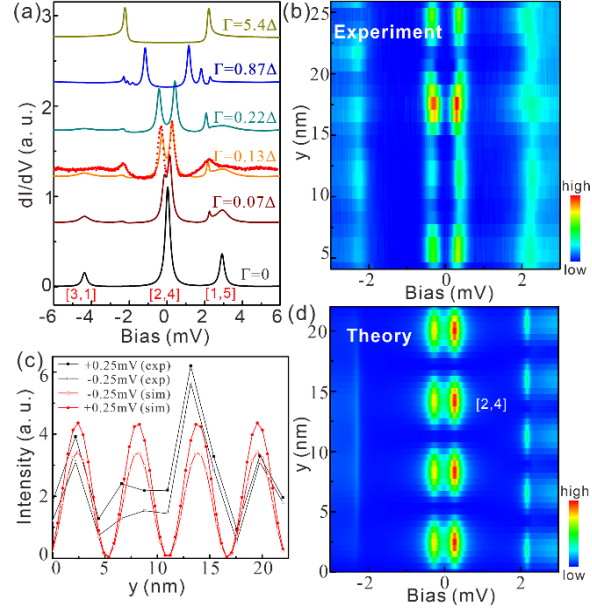

FIG. 3. (a) Calculated DOSs based on a model considering a QD proximity-coupled to a superconductor with different coupling strength Γ. QD's dimensions take the similar values of island #1 in Fig. 2(a). The black curve is the quantum levels simulated without considering the proximity effect, from which three discrete peaks near Fermi energy are observed (the corresponding quantum numbers [$n_x$, $n_y$] is given near each peak). The simulation in weakly coupling regime (orange curve), *e. g.* Γ = 0.13Δ, fits well to the experimental data (red circles) coming from Fig. 2(b). (b) A zoom-in plot of the spatially-resolved tunneling spectra from Fig. 2(c). (c) Position-dependence of the in-gap peak intensities (experimental data, black curves; simulations, red lines), which periodically oscillate with position in the island. (d) Calculated spatial distribution of in-gap states, which reproduces the key features, *i.e.* the number of nodes, of the data in (b).

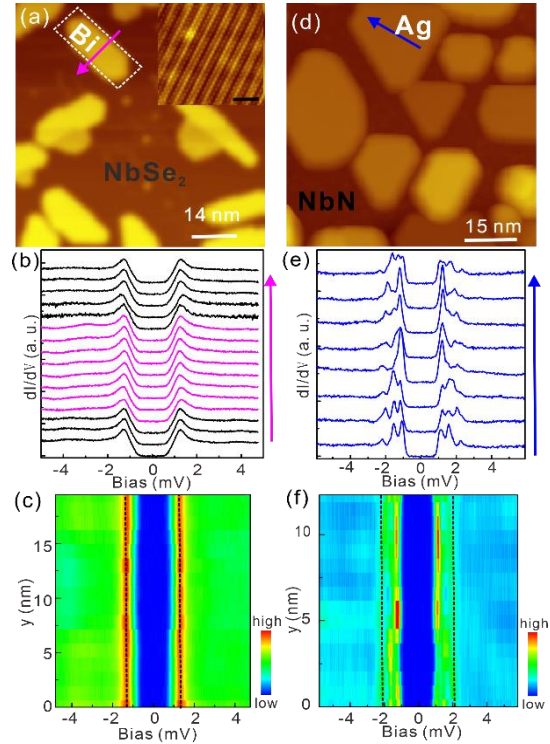

FIG. 4. (a) Topographic image of 1BL Bi(110) islands on NbSe$_2$. The inset: atomically resolved Bi(110) on NbSe$_2$. The scale bar is 2nm. (b) Spatially resolved tunnelling spectra (with offsets) measured along the red arrow in (a). Black (red) curves are measured on substrate (on island). No in-gap state appears. (c) Color plot of the position-dependent tunneling spectra shown in panel (b). The black dash lines mark the coherence peaks. (d) STM topography of Ag islands grown on NbN. (e) Spatially resolved tunnelling spectra (with offsets) taken along the blue arrow in (d). (f) Color plot of the position-dependent tunneling spectra shown in panel (e). Black dash lines indicates the coherence peaks. Spatially oscillated in-gap states are revealed.